\documentclass[aps,onecolumn,preprintnumbers,amsmath,amssymb]{revtex4}
\usepackage{graphicx}
\begin{document}

\title[]{Squeezing in multi-mode nonlinear optical state truncation}

\author{R.S. Said$^{\dag\ddag}$}
\email[Corresponding author\\E-mail address: ]{ressa.ss@mimos.my}

\author{M.R.B. Wahiddin$^{\dag\ddag}$}

\author{B.A. Umarov$^{\ddag}$}

\affiliation{$^{\dag}$Information Security Group, MIMOS Berhad, Technology Park Malaysia, 57000 Kuala Lumpur, Malaysia.\\$^{\ddag}$Faculty of Science, International Islamic University Malaysia (IIUM), Jalan Gombak, 53100 Kuala Lumpur, Malaysia.}

\begin{abstract}
In this paper, we show that multi-mode qubit states produced via nonlinear optical state truncation driven by classical external pumpings exhibit squeezing condition. We restrict our discussions to the two and three-mode cases.
\end{abstract}

%\preprint{PLA-D-06-02488 (Revised version)}

\maketitle

%%%%%%%%%%%%%%%%%%%%%%%%%%%%%%%%%%%%%%%%%%%%%%%%%%
% INTRODUCTION 																	 %
%%%%%%%%%%%%%%%%%%%%%%%%%%%%%%%%%%%%%%%%%%%%%%%%%%
\section{Introduction}

Squeezed state of light is one of the interesting phenomena that could only be properly explained by the quantum theory of light and has a lot of potential applications. For instance, it can be applied in quantum communications and measurements as well as novel spectroscopy with two-level atoms in squeezed fields. In terms of quantum communications, squeezed states are beneficial especially for reducing the noise in quantum channels \cite{yuen}. The basic ideas underlying squeezed states of light involve quantum noise or fluctuations in the quadrature components of the field and the uncertainty principle \cite{ridza}. It can be achieved in various forms involving non-linear optical interactions \cite{slusher,shelby,wu}. Prediction and observation of squeezing in quantum solitons has been done by Drummond et al. \cite{drummond}, and Rosenbluh and Shelby \cite{rosenbluh}. It was shown that squeezing in quantum solitons has also phase diffusion which occurs in 4-rank non-linear susceptibility media \cite{rosenbluh,carter,haus1}. A quantum non-demolition measurement of photon number can be performed by using collisions of quantum solitons \cite{haus2,watanabe,friberg}. Nonlinear optical techniques \cite{bloembergen} are the basis for developing experimental techniques to generate squeezing. Depending on the non-linearity and number of interacting modes, there are a variety of possible methods \cite{kimble}. Direct squeezing of propagating modes in a waveguide gives a relatively simple, broad-band implementation of quadrature squeezing \cite{slusher2,caves}. It is most often to get squeezing via four-wave mixing with short pulses or solitons because of availability of high-quality silica fiber \cite{rosenbluh,haus1,lai,drummond2}. Squeezing generation in the Kerr nonlinear coupler has been investigated theoretically and calculated numerically via a set of stochastic differential equations derived using positive P and Wigner representations \cite{basset}. This latter proposal motivates us to investigate the squeezing properties of multi-mode truncated states via optical state truncation or quantum scissors in a pumped non-linear coupler. 

Optical state truncation using quantum scissors was first proposed by Pegg et al. \cite{pegg}. The purpose is to truncate a single-mode coherent state of light to a superposition only of a vacuum and single-photon states. Later on, several papers are devoted to develop and generalize quantum scissors for optical state truncation based on linear media \cite{konio}. Experimental tests for quantum scissors for optical state truncation have already been reported in several reports \cite{babi}. In nonlinear regime, quantum scissors for single-mode optical state truncation can also be performed in Kerr media \cite{leo1}. A model of a non-linear coupler excited by a single-mode coherent field and filled with Kerr media was investigated \cite{miran} and can be considered as a two-mode nonlinear optical state truncation. The use of a conditional eight-port interferometry was proposed to implement optical-state truncation and teleportation of qudits \cite{miran2}. Nonlinear optical state truncation can also be experimentally feasible to generate the W state \cite{said}. 

In this paper, we study the quadrature variances of multi-mode truncated states generated by non-linear quantum scissors for optical state truncation with external classical pumping. We devote Section 2 to the case of truncated two-mode optical state while truncated three-mode state discussions are presented in Section 3. It is important here to inform the reader early that squeezing conditions can be observed for these two cases.

%%%%%%%%%%%%%%%%%%%%%%%%%%%%%%%%%%%%%%%%%%%%%%%%%%
% TRUNCATED TWO MODE  													 %
%%%%%%%%%%%%%%%%%%%%%%%%%%%%%%%%%%%%%%%%%%%%%%%%%%

\section{Truncated two-mode optical states}

We firstly provide in this section a brief review of important notions on two-mode optical state truncation in Kerr nonlinear coupler proposed by \cite{miran} and later define squeezing operators acting on the truncated states and investigate their quadrature variances.

One of the key ideas of two-mode nonlinear optical state truncations is to produce highly non-classical state of light i.e. two-qubit states by truncating two-mode optical state which can be written in the Fock representation of a time dependent wave function:
\begin{eqnarray}
|\psi(t)\rangle=\sum_{n,m=0}^{\infty}c_{nm}|n\rangle\otimes|m\rangle=\sum_{n,m=0}^{\infty}c_{nm}|n,m\rangle.
\end{eqnarray}
via a particular physical model satisfying the Hamiltonian below:
\begin{eqnarray}
\hat{H}_{tot1} = \hat{H}_{non1}+\hat{H}_{int1}+\hat{H}_{ext1}, \label{htot1}
\end{eqnarray}
where
\begin{eqnarray}
\hat{H}_{non1} &=& \frac{\chi_1}{2}(\hat{a}^\dagger_1 )^2\hat{a}^2_1+
\frac{\chi_2}{2}(\hat{a}^\dagger_2 )^2\hat{a}^2_2, \\
\hat{H}_{int1} &=& \epsilon \hat{a}_1^\dagger \hat{a}_2 + \epsilon^*\hat{a}_1\hat{a}_2^\dagger, \\
\hat{H}_{ext1} &=& \alpha_1\hat{a}_1^\dagger+\alpha_1^*\hat{a}_1+\alpha_2\hat{a}_2^\dagger+\alpha_2^*\hat{a}_2.
\end{eqnarray}
The total Hamiltonian of the system consists of nonlinear Hamiltonian $\hat{H}_{non1}$, interaction Hamiltonian $\hat{H}_{int1}$, and external pumping Hamiltonian $\hat{H}_{ext1}$. The system consists of two nonlinear oscillators coupled to one another in which the coupling strength is parameterized by $\epsilon$. The internal coupling is considered linear and the oscillators can be externally pumped by linear and constant amplitude coherent fields, $\alpha_1$ and $\alpha_2$.
$\chi_1$ and $\chi_2$ denote Kerr nonlinearity constants of each modes. 
$\hat{a}_1$ and $\hat{a}_2$ ($\hat{a}_1^\dag$ and $\hat{a}_2^\dag$) are photon annihilation (creation) operators acting on the mode 1 and 2 respectively. We use the assumption that the nonlinearity constants are much greater than all coupling parameters. Straightforwardly, the state evolution can be viewed as a resonant case in which some subspace of the wave function have a small probability. 
For the system with single pumping, i.e. $\alpha_2=0$, we can get the truncated state as: 
\begin{eqnarray}
|\psi_{tr}(t)\rangle &=& \sum_{n,m=0}^{1}c_{nm}|n,m\rangle =c_{00}(t) |00\rangle + c_{01}(t) |01\rangle + c_{10}(t) |10\rangle + c_{11}(t) |11\rangle, \label{Psitr1}
\end{eqnarray}
while $c_{nm}$ is a complex amplitude which gives probability of finding the system in the $n$-photon and $m$-photon modes. Analytical probability amplitudes $c_{nm}$ for $n,m\in \{0,1\}$, which are given by applying the rotating wave approximation (RWA)-like approach to a set of equations for the amplitudes obtained from the Schrodinger equation, are simply written as:
\begin{eqnarray}
c_{00}(t)&=&\cos{(x_1 t)}\cos{(y_1 t)} + (1/\sqrt{5}) \sin{(x_1 t)} \sin{(y_1 t)}, \:
c_{01}(t)=-(2/\sqrt{5}) \sin{(x_1 t)} \sin{(y_1 t)}, \cr
c_{10}(t)&=&-i(2/\sqrt{5})\cos{(x_1 t)} \sin{(y_1 t)}, \:
c_{11}(t)= i \left( (1/\sqrt{5}) \cos{(x_1 t)}\sin{(y_1 t)} + \sin{(x_1 t)} \cos{(y_1 t)} \right) \label{quadrature2},
\end{eqnarray}
where $x_1=\alpha_1/1$ and $y_1=\sqrt{5}x_1$. $\epsilon$ and $\alpha_1$ are chosen to be real and equal. Here, we set our initial condition such that $c_{00}(t=0)=1$ and take $\hbar=1$. For the system with two external pumpings \cite{miran3}, we have the amplitudes of the equation (\ref{Psitr1}) written as:
\begin{eqnarray}
c_{00} &=& 1/2 \left( 1+ \left( \cos (\lambda t/2)+ i (\epsilon/\lambda) \sin (\lambda
t/2)\right) e^{-i\epsilon t/2} \right), \nonumber \\
c_{01} &=& c_{10} = -(2i\alpha/\lambda)  \sin (\lambda
t/2) e^{-i\epsilon t/2}, \:
c_{11} = c_{00}-1, \label{amp2}
\end{eqnarray}
where $\alpha_1=\alpha_2=\alpha$ and $\lambda=\sqrt{16\alpha^2+\epsilon^2}$. All coupling parameters taken above are assumed to be real and positive.

In order to investigate the squeezing phenomena in the system, we now define the quadrature variances for each mode $p={1,2}$, of the truncated states as follow:
\begin{eqnarray}
\langle \Delta \hat{X}^2_{p} \rangle=\langle \hat{X}^2_{p} \rangle - \langle \hat{X}_{p} \rangle ^2 = \langle
\psi_{tr}(t)| \hat{X}_{p} \hat{X}_{p} |\psi_{tr}(t)\rangle - \langle \psi_{tr}(t) | \hat{X}_{p} | \psi_{tr}(t)\rangle ^2 \label{variance1} 
\end{eqnarray}
and
\begin{eqnarray}
\langle \Delta \hat{Y}^2_{p} \rangle= \langle \hat{Y}^2_{p} \rangle - \langle \hat{Y}_{p} \rangle ^2 = \langle
\psi_{tr}(t) | \hat{Y}_{p} \hat{Y}_{p} |\psi_{tr}(t)\rangle - \langle \psi_{tr}(t) | \hat{Y}_{p} | \psi_{tr}(t) \rangle ^2, \label{variance2}
\end{eqnarray} 
where the Hermitian operators in the two equations above, $\hat{X}_p$ and $\hat{Y}_p$, can be properly described as
\begin{eqnarray}
\hat{X}_{p}=\frac{1}{2} \left( \hat{a}_p+\hat{a}_p^{\dag} \right) \label{observable1},\quad 
\hat{Y}_{p}=\frac{1}{2i} \left( \hat{a}_p-\hat{a}_p^{\dag} \right) \label{observable2}. 
\end{eqnarray}
It is then obvious that the squeezing condition for the above operator is $\langle \Delta \hat{X}^2_{p} \rangle < \frac{1}{4}$ or $\langle \Delta \hat{Y}^2_{p}\rangle < \frac{1}{4}$.
For the case of the system with single pumping, the two quadrature variances for each modes can be analytically described as:
\begin{eqnarray}
\langle \Delta \hat{X}^2_{1} \rangle &=& -(\sqrt{5}/5){\gamma_1}{\gamma_2}{\phi_1}{\phi_2} + 1/4\left(\left( 2\phi_1^2 +1 \right)\gamma_2^2+ \left( 2\gamma_1^2+ 1 \right)\phi_2^2 \right), \nonumber \\
\langle \Delta \hat{X}^2_{2} \rangle &=& -4\phi_1^4 /25  -\gamma_1/5 \left( 4 \sqrt{5} \gamma_2 \phi_1 \phi_2 - \left( 2 \phi_2^2 + 1 \right) 5 \gamma_1 \right) +\phi_1^2/20 \left( 6 \phi_2^2 + 7 \right),\nonumber\\
\langle \Delta \hat{Y}^2_{1} \rangle &=& 1/4 \left( \left( 3-2\gamma_2^2 \right)\gamma_1^2 + \left(3-2\phi_2^2\right) \phi_1^2 \right) + 16/25
\left( \sqrt{5}\left( 2\phi_2^2 - 1\right) \gamma_1- \gamma_2\phi_1\phi_2\right)\gamma_2\phi_1^3\phi_2\nonumber \\&&+
1/5 \left( \left(-\sqrt{5}+8\gamma_1\gamma_2\phi_1\phi_2\right)\gamma_1\gamma_2\phi_1\phi_2
-4\left(\gamma_2^4+\phi_2^4\right)\phi_1^2\gamma_1^2\right), \nonumber\\
\langle \Delta \hat{Y}^2_{2} \rangle &=& 1/20 \left( \left(5 +2 \phi_1^2 \right) \gamma_2^2 +
13 \phi_1^2 \phi_2^2 - \gamma_1 \phi_2 \left(4 \sqrt{5} \gamma_2 \phi_1  - 15 \gamma_1 \phi_2 \right)\right),
\end{eqnarray}
where $\gamma_1=\cos\left(\frac{1}{2}\sqrt{5}\alpha_1 t\right)$, $\gamma_2=\cos\left(\frac{1}{2}\alpha_1 t\right)$, $\phi_1=\sin\left(\frac{1}{2}\sqrt{5}\alpha_1 t\right)$ and $\phi_2=\sin\left(\frac{1}{2}\alpha_1 t\right)$. 
We can also obtain analytical expressions of quadrature variances for the case of two pumpings as follow:
\begin{eqnarray}
\langle \Delta \hat{X}^2_{1} \rangle &=& 
\Xi_1 \cos^2\tau_1
+\Xi_2\sin\tau_1\sin\tau_2\cos\tau_1
+\Xi_3
+\Xi_4 \cos^4\tau_1
+\left(\epsilon^2/2\lambda^2 \right) \cos 2 \tau_2\sin^22\tau_1-(1/4)\cos\tau_1 \nonumber\\
&&
+\sin\tau_2\sin\tau_1\left(\left(2\epsilon/\lambda\right)^3 \cos^3\tau_1-(\epsilon/4\lambda)\right),\nonumber\\
\langle \Delta \hat{X}^2_{2} \rangle &=&
\Xi_5\cos^4\tau_1 -\Xi_6\sin\tau_1\sin\tau_2\cos\tau_1 +\Xi_7
+\Xi_8\cos^2\tau_1 +\left(\epsilon^2/2\lambda^2 \right) \cos 2 \tau_2\sin^22\tau_1\nonumber\\
&&+\left(2\epsilon/\lambda\right)^3\sin\tau_2\sin\tau_1\cos^3\tau_1, \cr
\langle \Delta \hat{Y}^2_{1} \rangle &=&
\Xi_9\cos^2\tau_1+\Xi_{10}-2\left(\epsilon/\lambda\right)^2\cos2\tau_2\sin^2\tau_1
-\left(\epsilon/2\lambda\right)\sin\tau_2\sin\tau_1\left(1/2-\cos\tau_1\right) -1/4\cos\tau_1  
, \cr
\langle \Delta \hat{Y}^2_{2} \rangle &=&
\Xi_{11}\cos^2\tau_1 + \Xi_{12}
- \left(2/\lambda^2\right)\cos2\tau_2\left(\epsilon^2 - \cos^2\tau_1\right) +\left(\epsilon/4\lambda\right)\sin\tau_2\sin\tau_1\cos\tau_1.
\end{eqnarray}
We here use $\tau_1=\lambda t/2$ and $\tau_2=\epsilon t/2$ while $\Xi_i(i={1...12})$ are given by: 
\begin{eqnarray}
\Xi_1&=&(\lambda^4+32\epsilon^4-25\lambda^2\epsilon^2)/4\lambda^4,
\: \Xi_2=(\lambda^2\epsilon-16\epsilon^3)/2\lambda^4, \:
\Xi_3=(17\lambda^2\epsilon^2+\lambda^4-16\epsilon^4)/4\lambda^4,
\cr \Xi_4&=&(8\lambda^2\epsilon^2-16\epsilon^4)/4\epsilon^4, \: 
\Xi_5=2(\lambda^2\epsilon^2-\epsilon^4)/\lambda^4,
\: \Xi_6=8\epsilon(\epsilon^2-\lambda^2/32)/\lambda^3, \:
\Xi_7=(49\lambda^2\epsilon^2+\lambda^4-32\epsilon^4)/8\lambda^4,
\cr \Xi_8&=&(64\epsilon^4+\lambda^4-65\lambda^2)/8\lambda^4, \:
\Xi_{9,10}=(\lambda^2 \mp 9\epsilon^2)/4\lambda^2,
\: \Xi_{11,12}=(\lambda^2 \mp 33\epsilon^2)/8\lambda^2.
\end{eqnarray}

\begin{figure}[t]
	\centering
	%\begin{tabular}{cc}
		\includegraphics[scale=0.6]{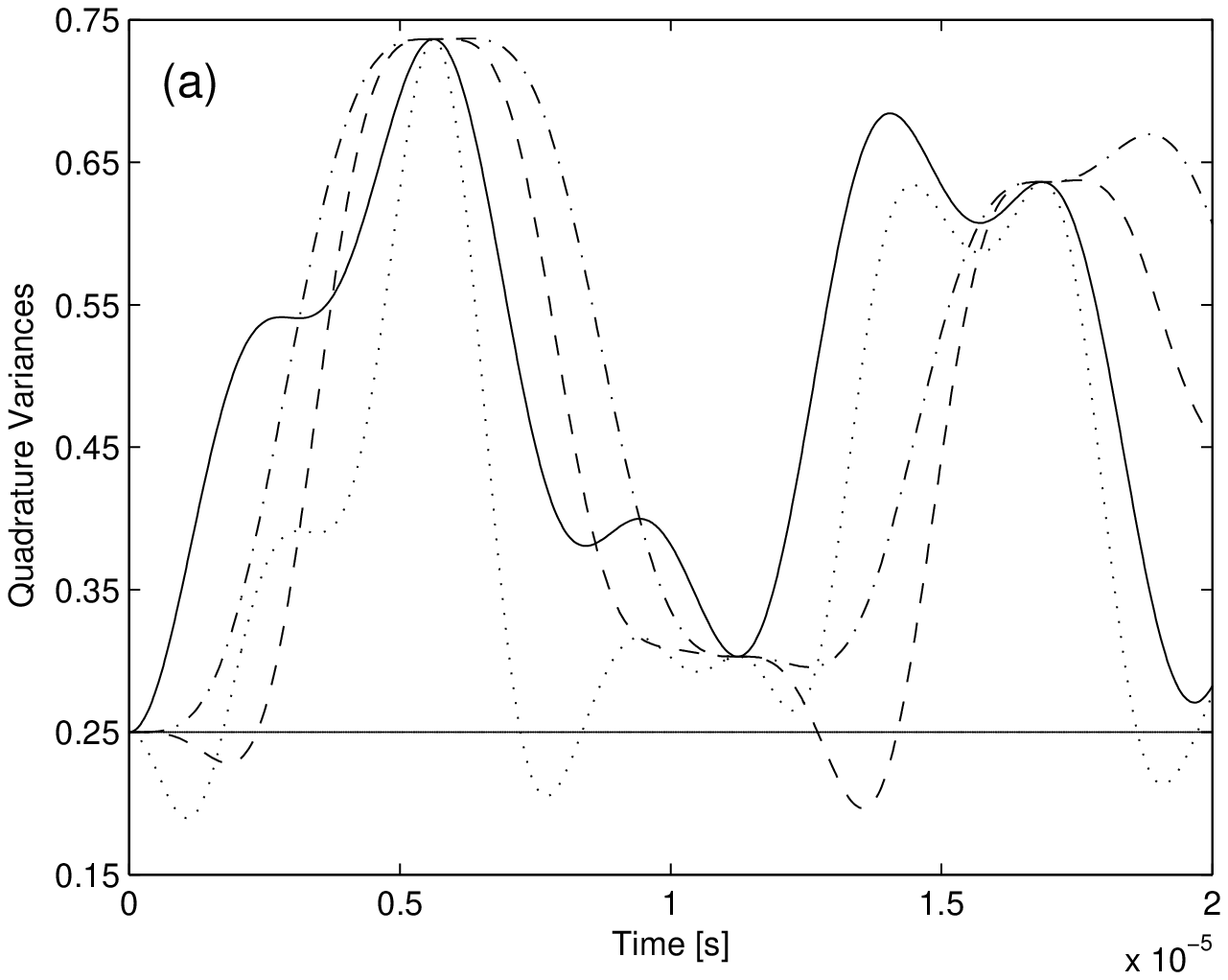}
		%&
		\includegraphics[scale=0.6]{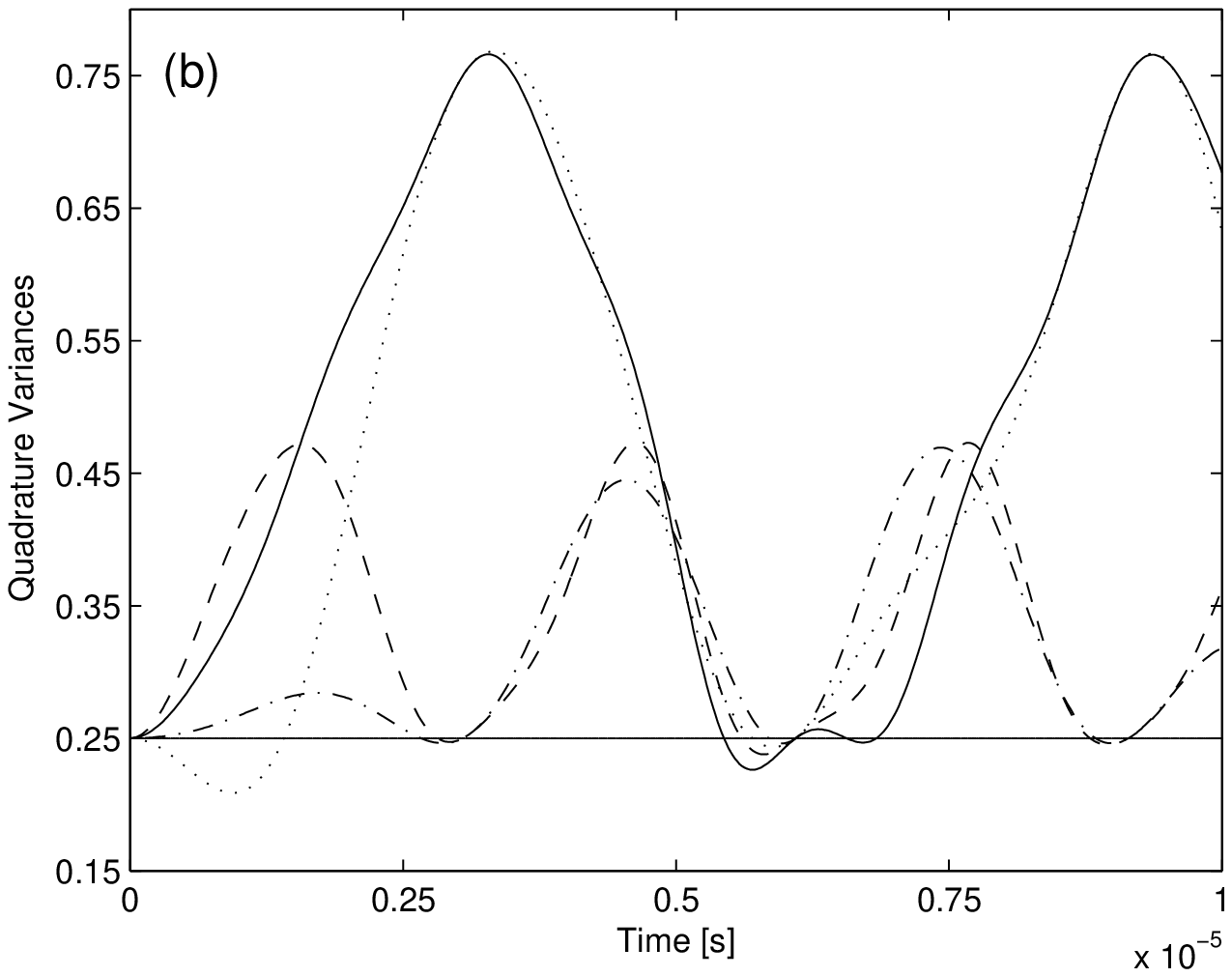}
	%\end{tabular}
		\caption{The evolution of quadrature variances of two-mode truncated state for the system with single pumping {\bf(a)} and two pumpings {\bf(b)}: $\langle \Delta \hat{X}^2_{1} \rangle$ (solid line), $\langle \Delta \hat{X}^2_{2} \rangle$ (dashed line), $\langle \Delta \hat{Y}^2_{1} \rangle$ (dotted line) and $\langle \Delta \hat{Y}^2_{2} \rangle$ (dash-dot line). In calculations we take the coupling strength parameters, $\epsilon=\alpha_1=\alpha_2=10^8/200$.}\label{fig1}
\end{figure} 
In Figure \ref{fig1}, we show that the quadrature variances oscillate as a function of time for the system with single and two pumpings. Here we use $\epsilon=\alpha_1=10^8/200$, following the parameter values as used in \cite{miran3}. For $t\leq 5 \mu s$, the first squeezing condition of the truncated state produced from the system with single pumping is satisfied for the observable $\hat{X}$ on mode 2 at the time interval of $0.06\mu s<t<2.37\mu s$ and $\hat{Y}$ on mode 1 at the time interval of $0<t<1.74\mu s$. Local minima (i.e. maximum squeezing) of $\hat{X}_{2}$ at $t\leq 5 \mu s$ occurs at $t\approx 1.83\mu s$ while for $\hat{Y}_{1}$ is at $t\approx 1.08\mu s$.
In the case of two external pumpings, the observable $\hat{X}$ acting on the mode 1 satisfies the squeezing condition for the time interval $5.42\mu s<t<6.10\mu s$ and $6.56\mu s<t<6.84\mu s$ while for mode 2 the time interval needed is $2.80\mu s<t<3.06\mu s$, $5.58\mu s<t<6.10\mu s$, and $8.78\mu s<t<9.16\mu s$. Here, we use the restriction of the time interval $t<10\mu s$. Maximum squeezing condition for the observable $\hat{X}$ appeared for the mode 1 at $t\approx 5.70\mu s$. As shown in Figure \ref{fig1}(b), during the time interval of $0<t<1.42\mu s$ and $5.68\mu s<t<6.10\mu s$, the squeezing conditions for mode 1 are fulfilled by the observable $\hat{Y}$. On the other mode, i.e. mode 2, the condition $\langle \Delta \hat{Y}^2_{2} \rangle < \frac{1}{4}$ is met for $2.66\mu s<t<3.06\mu s$, $5.82\mu s<t<6.10\mu s$ and $8.84\mu s<t<9.16\mu s$. The observable $\hat{Y}$ gives a minimum quadrature variance on mode 1 at $t\approx 0.94\mu s$.

Squeezing conditions appear only when the system is attached with external classical pumpings such that the existence of external coupling parameters $\alpha_i(i=1,2)$ play an important role in generating such conditions. In order to see the validity of the statement we can then consider the case of the system without pumpings i.e. by removing the term of $H_{ext1}$ in the equation (\ref{htot1}). In this case we have a situation where energy of the system is conserved. The truncated states produced from the system without pumping can be written as
$
|\psi_{tr}(t)\rangle_0 = -i\sin \epsilon t |01\rangle + \cos \epsilon t |10\rangle.
$
Hence, we can write the equations for quadrature variances of the state $|\psi_{tr}(t)\rangle_0$ as follow
\begin{eqnarray}
\langle \Delta \hat{X}^2_{1} \rangle_{0} = \langle \Delta \hat{Y}^2_{1} \rangle_{0} = 1/4\left(1+2\cos^2 \epsilon t \right), \:
\langle \Delta \hat{X}^2_{2} \rangle_{0} = \langle \Delta \hat{Y}^2_{2} \rangle_{0} = 1/4\left(1+2\sin^2 \epsilon t \right). \label{nopump}
\end{eqnarray}
We clearly see here that we can never observe squeezing condition in the system without external pumping. It then justifies the necessity for the external classical field fed to the system.

%%%%%%%%%%%%%%%%%%%%%%%%%%%%%%%%%%%%%%%%%%%%%%%%%%
% THREE MODE  																	 %
%%%%%%%%%%%%%%%%%%%%%%%%%%%%%%%%%%%%%%%%%%%%%%%%%%

\section{Truncated three-mode optical states}

We next generalize the idea of optical state truncation to a three mode optical state which can obviously be expressed in the Fock representation as follows:
\begin{eqnarray} \label{zeta}
|\zeta(t)\rangle=\sum_{u,v,w=0}^{\infty}c_{uvw}|u,v,w\rangle,
\end{eqnarray}
where we use the same definition as before but now the amplitude, $c_{uvw}$ is a complex probability of finding the state in the $u$-photon, $v$-photon, and $w$-photon bases for each mode.
This generalization was firstly highlighted in \cite{said} in the context of tripartite non maximal entanglement generation i.e. the W state. We require an extension of the physical model proposed by Leonski and Miranowicz \cite{miran},
three nonlinear oscillators are now coupled to each other \cite{said}. Taking the same assumptions used in the previous section, one can optionally drive each oscillator by external electromagnetic field. We here again consider that the field has constant amplitude and is excited linearly. In the feasible implementation regime, a high-{\it Q} cavity is strongly demanded because we have to preserve the radiation such that we could approach a condition of neglected damping process \cite{highQcavity}. We can write the total Hamiltonian $\hat{H}_{tot2}$, of the proposed system as:
\begin{eqnarray} \label{Htot2}
\hat{H}_{tot2} &=& \hat{H}_{non2}+\hat{H}_{int2}+\hat{H}_{ext2},
\end{eqnarray}
where
\begin{eqnarray} \label{Htot2b}
\hat{H}_{non2} &=& \frac{\chi_1}{2}(\hat{a}^\dagger_1 )^2\hat{a}^2_1+
\frac{\chi_2}{2}(\hat{a}^\dagger_2 )^2\hat{a}^2_2 + \frac{\chi_3}{2}(\hat{a}^\dagger_3 )^2\hat{a}^2_3 \\
\hat{H}_{int2} &=& \epsilon \hat{a}_1^\dagger \hat{a}_2 + \epsilon^*\hat{a}_1\hat{a}_2^\dagger 
+ \epsilon \hat{a}_1^\dagger \hat{a}_3 + \epsilon^*\hat{a}_1\hat{a}_3^\dagger
+ \epsilon \hat{a}_2^\dagger \hat{a}_3 + \epsilon^*\hat{a}_2\hat{a}_3^\dagger
\\
\hat{H}_{ext2} &=& \alpha_1\hat{a}_1^\dagger+\alpha_1^*\hat{a}_1+\alpha_2\hat{a}_2^\dagger+\alpha_2^*\hat{a}_2
+\alpha_3\hat{a}_3^\dagger+\alpha_3^*\hat{a}_3.
\end{eqnarray}
Applying Schr\"{o}dinger equation to the equations (\ref{Htot2}) and (\ref{zeta}) one will get a coupled set of the equations of motion for the amplitude in time domain and again for the sake of simplicity, we take $\hbar=1$.
\begin{eqnarray} \label{3mode}
i\frac{d}{dt}c_{u,v,w}&=&\frac{\chi_1}{2}c_{u,v,w}(u-1)u+\frac{\chi_3}{2}c_{u,v,w}(v-1)v+\frac{\chi_3}{2}c_{u,v,w}(w-1)w \nonumber
	\\&&+\epsilon c_{u-1,v+1,w}\sqrt{u}\sqrt{v+1}+\epsilon^* c_{u+1,v-1,w}\sqrt{u+1}\sqrt{v}
	+\epsilon c_{u-1,v,w+1}\sqrt{u}\sqrt{w+1}\cr &&+\epsilon^* c_{u+1,v,w-1}\sqrt{u+1}\sqrt{w}
	+\epsilon c_{u,v-1,w+1}\sqrt{v}\sqrt{w+1}+\epsilon^* c_{u,v+1,w-1}\sqrt{v+1}\sqrt{w}\nonumber
	\\&&+\alpha_1 c_{u-1,v,w}\sqrt{u}+\alpha_1^* c_{u+1,v,w}\sqrt{u+1}
	+\alpha_2 c_{u,v-1,w}\sqrt{v}\nonumber \\&&+\alpha_2^* c_{u,v+1,w}\sqrt{v+1}
	+\alpha_3 c_{u,v,w-1}\sqrt{w}+\alpha_3^* c_{u,v,w+1}\sqrt{w+1}.
\end{eqnarray}
The dynamics of the system can be closed and some subspaces of states in Fock state representation has a negligible probability by assuming all coupling parameters are weak compared to the nonlinearity i.e. $\epsilon,\alpha_i << \chi_i$ for $i=1,2,3$. The transition of the state evolved can be treated as a resonant case and using the same approximation methods as that used in previous section, we can neglect the influence of the probability amplitude for $u,v,w\leq 2$. We can analytically solve the equation (\ref{3mode}) by assuming the system as symmetrical i.e. all coupling parameters are real and $\alpha_1=\alpha_2=\alpha_3=\epsilon$ \cite{said} such that our wave function now becomes
\begin{eqnarray} \label{zetatr}
|\zeta_{tr}(t)\rangle=\sum_{u,v,w=0}^{1}c_{uvw}|u,v,w\rangle,
\end{eqnarray}
where the amplitudes are given by:
\begin{eqnarray} \label{amplitude}
c_{000}&=&e^{-2i\epsilon t} \left( (\sqrt{7}/7) i \sin \sqrt{7} \epsilon t 
+ (1/2) \cos \sqrt{7} \epsilon t \right) + (1/2) \cos \sqrt{3} \epsilon t, \nonumber\\
c_{001}&=&-\sqrt{7}/14 \left(i \cos 2 \epsilon t \sin \sqrt{7} \epsilon t +
\sin 2 \epsilon t \sin \sqrt{7} \epsilon t \right) - (\sqrt{3}/6) i \sin \sqrt{3} \epsilon t, \nonumber\\
c_{010}&=&c_{100}=c_{001}, \:
c_{011}=c_{101}=c_{110}=c_{001} + (\sqrt{3}/3) i \sin \sqrt{3} \epsilon t, \:
c_{111}=c_{000} - \cos \sqrt{3} \epsilon t.
\end{eqnarray}
In asymmetrical cases, i.e. the coupling parameters are not exactly same, the equation (\ref{3mode}) is solved numerically. The detail of calculations and results with regards to complex amplitudes in asymmetrical cases is presented in \cite{thesis}. Apart from the symmetrical system with same external coupling strength, we here only consider the case of single pumping with $\alpha_1=\epsilon$ and two pumpings with $\alpha_1=\alpha_2=\epsilon$.

To investigate the squeezing, we need to extend equations (\ref{variance1})-(\ref{observable2}) i.e. $p=1,2,3$ such that our quadrature variances become $\langle \Delta \hat{X}^2_{1} \rangle$, $\langle \Delta \hat{X}^2_{2} \rangle$, $\langle \Delta \hat{X}^2_{3} \rangle$, $\langle \Delta \hat{Y}^2_{1} \rangle$, $\langle \Delta \hat{Y}^2_{2} \rangle$, and $\langle \Delta \hat{Y}^2_{3} \rangle$. Figures 2 and 3 present numerical simulations of the evolution of quadrature variances of three mode truncated states with single, two and three pumpings respectively. We here consider the evolution of quadrature variances only up to time $t=2\times10^{-5}s$.
\begin{figure}[t]
	\centering
		%\begin{tabular}{cc}
		\includegraphics%[width=221pt,height=195pt]
		[scale=0.6]
		{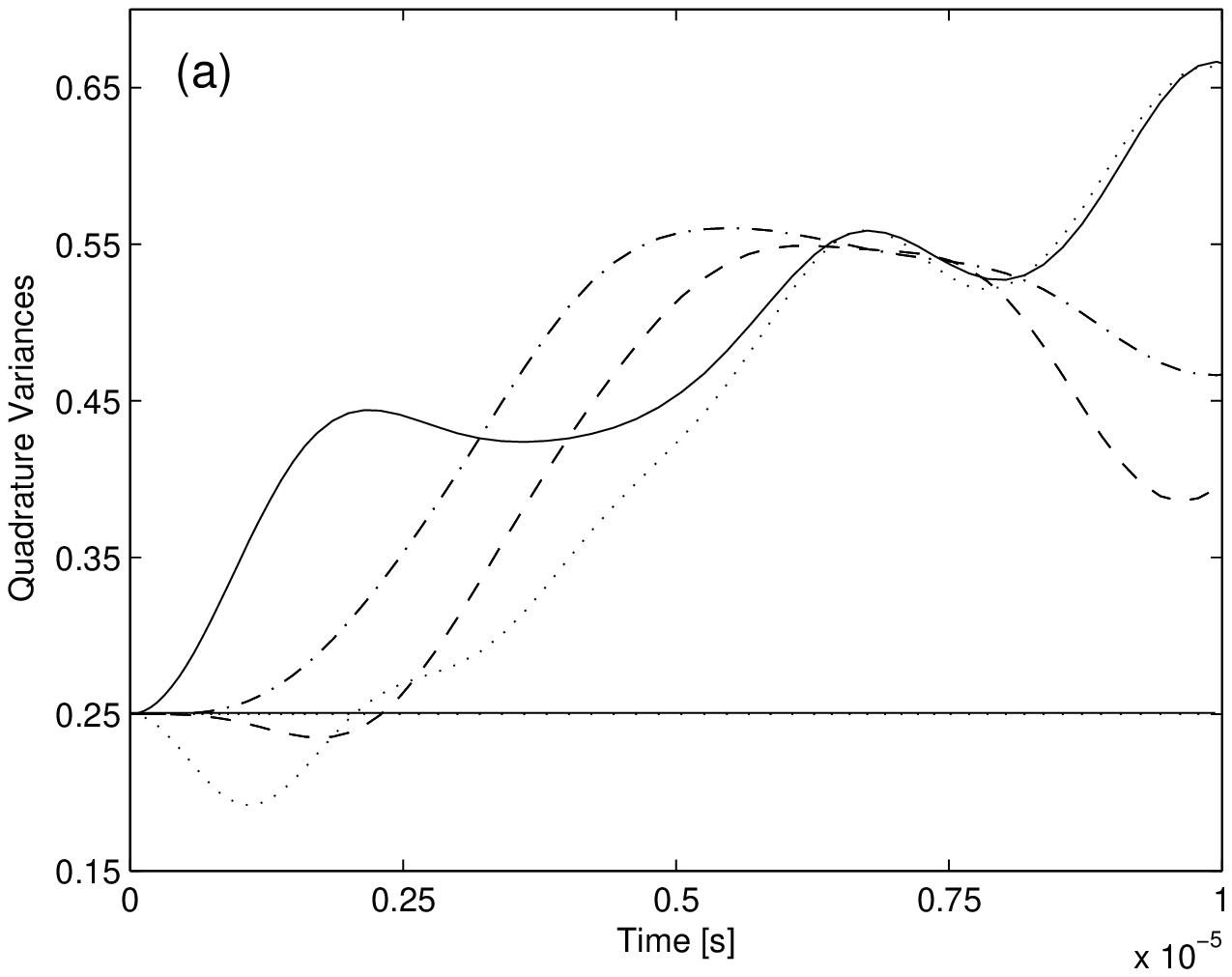}
		%&
		\includegraphics[scale=0.6]{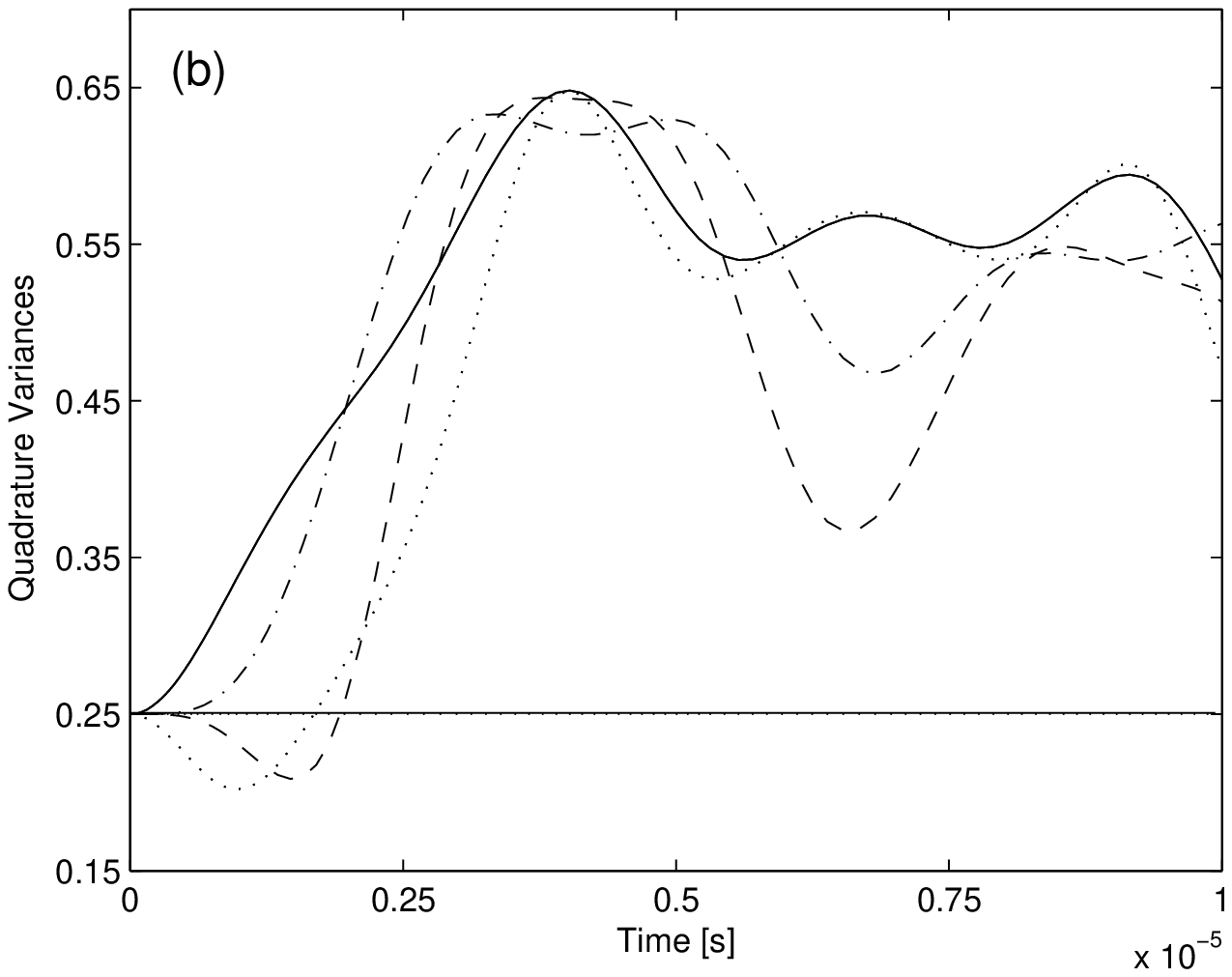}
		%\end{tabular}
	\label{fig45}
	\caption{{\bf(a)} The evolution of quadrature variances of three-mode truncated state for the system with single pumping: $\langle \Delta \hat{X}^2_{1} \rangle$ (solid line), $\langle \Delta \hat{X}^2_{2} \rangle=\langle \Delta \hat{X}^2_{3} \rangle$ (dashed line), $\langle \Delta \hat{Y}^2_{1} \rangle$ (dotted line) and $\langle \Delta \hat{Y}^2_{2} \rangle$=$\langle \Delta \hat{Y}^2_{3} \rangle$ (dash-dot line). The coupling strength parameters are $\epsilon=\alpha_1=10^8/200$. {\bf(b)} The evolution of quadrature variances of three-mode truncated state for the system with two pumpings: $\langle \Delta \hat{X}^2_{1} \rangle=\langle \Delta \hat{X}^2_{2} \rangle$ (solid line) $\langle \Delta \hat{X}^2_{3} \rangle$ (dashed line), $\langle \Delta \hat{Y}^2_{1} \rangle =\langle \Delta \hat{Y}^2_{2} \rangle$ (dotted line) and $\langle \Delta \hat{Y}^2_{3} \rangle$ (dash-dot line). The coupling strength parameters are $\epsilon=\alpha_1=\alpha_2=10^8/200$.}
\end{figure}
\begin{figure}[t]
	\centering
		\includegraphics[scale=0.6]{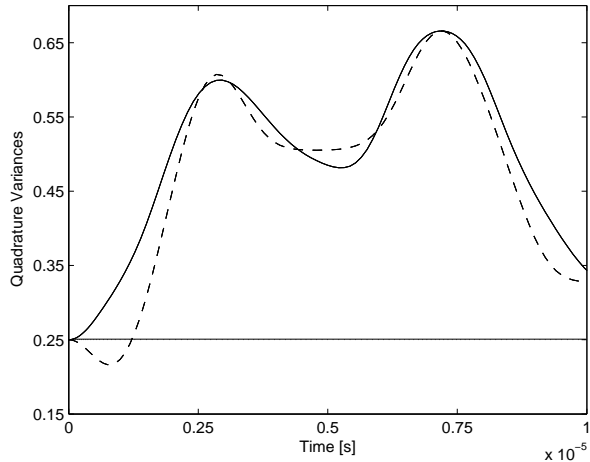}
	\label{fig6}
	\caption{The evolution of quadrature variances of three-mode truncated state for the system with three pumpings: $\langle \Delta \hat{X}^2_{1} \rangle=\langle \Delta \hat{X}^2_{2} \rangle=\langle \Delta \hat{X}^2_{3} \rangle$ (solid line) and $\langle \Delta \hat{Y}^2_{1} \rangle=\langle \Delta \hat{Y}^2_{2} \rangle=\langle \Delta \hat{Y}^2_{3} \rangle$ (dashed line). The coupling strength parameters are $\epsilon=\alpha_1=\alpha_2=\alpha_3=10^8/200$.}
\end{figure}
 
Whilst the system is driven by single pumping, the quadrature variances of the two observables $\hat{X}$ and $\hat{Y}$ acting on mode 2 is the same as that which acts on mode 3. In this case, we observe the squeezing condition at certain times as shown by Figure 2(a) for observable $\hat{X}$ on modes 2 and 3 while the quadrature variance for observable $\hat{Y}$ satisfy the squeezing condition for mode 1. If two oscillators in the system are externally driven by classical pumping i.e. two pumpings, as depicted by Figure 2(b), the squeezing condition would occur in mode 3 for the observable $\hat{X}$ and in modes 1 and 2 for $\hat{Y}$. For the system driven by three pumpings, we see in Figure 3 that the observable $\hat{Y}$ is squeezed for every mode for some interval of time. It is important to note that although it is possible to come up with analytical expressions on the time evolution of quadrature variances for the system with symmetrical pumpings, we present none of them here in order to avoid some rigor and complicated mathematical expressions. Instead, we straightforwardly plot the quadrature variances.
%%%%%%%%%%%%%%%%%%%%%%%%%%%%%%%%%%%%%%
\section{Conclusions}
We have calculated analytically time evolutions of quadrature variances of two qubit states generated from nonlinear optical state truncation system pumped externally by classical fields and shown that these states exhibit squeezing conditions for some interval of time. We also investigate numerically the quadrature variance for the case of generalized system generating three-mode qubit state. It has been shown that the latter case could also satisfy squeezing condition at certain interval time.
It is necessary to mention here that we have considered the ideal system such that the experimental observation of such discussed behavior is still considered difficult to realize. However, extension of our work to the more practical system by considering damping process is not impossible and would open more insight to the subject.

%%%%%%%%%%%%%%%%%%%%%%%%%%%%%%%%%%%%
\section*{Acknowledgements}
%%%%%%%%%%%%%%%%%%%%%%%%%%%%%%%%%%%%%%
The authors are grateful to A Messikh for his constructive comments. 
R.S. Said thanks MIMOS Berhad for financial support.
This research was also supported by the Malaysia IRPA Grant 09-02-08-0203-EA002. 

%%%%%%%%%%%%%%%%%%%%%%%%%%%%%%%%%%%%%

\end{document}